\documentstyle[preprint,aps]{revtex}
\begin{document}
 \draft
 \title{Shape of domains in two-dimensional systems: \\virtual singularities
and
 a generalized Wulff construction}
\author{Joseph Rudnick and Robijn Bruinsma}
\address{Department of Physics\\UCLA\\Los Angeles, California 90024-1547}
\date{\today}
\maketitle
\begin{abstract}

We report on a generalized Wulff construction that allows for the
calculation of the shape of two-dimensional materials with orientational
order but no positional order.  We demonstrate that for sufficiently large
domain radii, the shape necessarily develops mathematical singularities,
similar to those recently observed in Langmuir monolayers.  The physical
origin of the cusps is shown to be related to the softness of the material
and is fundamentally diffferent from that of the sharp angles seen in the
shape of hard crystals.
\end{abstract}

The singularities in the shape of crystalline materials have long been
 understood to be a macroscopic expression of the positional order of crystals
 at the atomic level. The angles between the faces of a crystal are, for
 instance, structural invariants dependent only on a certain set of integers
 (Miller indices\cite{Miller}). In a classic paper\cite{Wulff}, Wulff developed
 a geometrical construction that allows one to determine the equilibrium shape
 of a crystal, provided one knows the values of the surface energies for the
 various Miller indices. If, for every index, the corresponding surface is
 placed a distance from a fixed point proportional to the surface energy, then
 the inner envelope of the planes constitutes the minimum energy crystal shape.

Surprisingly, sharp edges in the shapes of samples are not just encountered for
 hard, crystalline materials. Using the Wulff construction, Herring argued that
 liquid crystals without positional order, but with orientational order could
 also display sharp ables, or cusps\cite{HerringWulff}. The cusps are no longer
 material invariants. Cusped domain shapes are, indeed, well-documented for
 three dimensional liquid crystals\cite{3dcusp,OM}. More recently, studies of
 the shapes of two dimensional materials without positional order have also
 revealed cusps and sharp angles\cite{2dcusp}. These studies have also revealed
 that the internal structure of such ``soft'' materials is significantly
 deformed and dependent on the domain shape, while Herring assumed a rigid
 internal structure. Thus, the Wulff construction is not manifestly valid in
 this case.

In this paper we present a formalism which allows for the construction of
domain
 shapes of soft two dimensional materials when the internal strucure is
 described by a two dimensional $XY$ model (e. g. hexatic or nematic liquid
 crystals). We will demonstrate that cusps are, indeed, not just possible, but
 that they ought to be a {\em generic} feature of such domain shapes. Moreover,
 the cusp angle provides important information on the elastic moduli and
surface
 energy of the material.

We will model the internal structure of a domain by a unit vector
 $\hat{c}=\left(\cos\Theta(\vec{r}),\sin \Theta(\vec{r}) \right)$. The
 associated $XY$ model free energy is
\begin{equation}
{\rm H} \left[ \Theta (x,y) \right] = \int\frac{\kappa}{2}\left|\vec{\nabla}
\Theta \right|^2\,dx\,dy+\int_{ {\rm boundary}}ds\, \sigma\left(\theta-\Theta
 \right).
\label{1}
\end{equation}
Here, $\kappa$ is the stiffness (or Frank constant) of the order parameter
field
 and $\sigma$ is the anisotropic surface energy, which depends on the  relative
 angle, $\Theta -\theta$, between the order parameter and the outward normal
 $\hat{n}=\left(\cos \theta,\sin \theta \right)$ to the domain boundary. Note
 that $\sigma(\phi-2 \pi) = \sigma (\phi)$.

To find the optimal domain shape we must minimize ${\rm
 H}\left[\Theta(x,y)\right]$ with respect to $\Theta(x,y)$ and the domain
shape,
 while keeping the domain area, $A$, fixed. We start with the limiting case
 $\kappa = \infty$ when the texture is rigid so that $\Theta = 0$ and only the
 shape needs to be varied.

\begin{center}
\underline{ $\kappa=\infty$}
\end{center}

The domain shape is a closed planar curve. We will express the co-ordinates
 $(x,y)$ of a point on the curve in terms of the angle $\theta$ of $\hat{n}$
and
 the minimum distance $R(\theta)$ between the tangent line through $(x,y)$ and
 the origin. In terms of $R(\theta)$:
\begin{mathletters}
\label{2}
\begin{eqnarray}
x(\theta)&=&R(\theta)\cos(\theta)-\frac{dR(\theta)}{d\theta}\sin(\theta)
 \\
y(\theta)&=&R(\theta)\sin(\theta)+\frac{dR(\theta)}{d\theta}\cos(\theta)
{}.
\end{eqnarray}
\end{mathletters}
The variational equation determining the domain shape is, then
\begin{equation}
\frac{\delta}{\delta R(\theta)}
\oint\left[\sigma(\theta')
\left[R(\theta')+\frac{d^2R(\theta')}{d\theta'^2}\right] -\frac{
\lambda} {2} R(\theta')
\left[R(\theta')+\frac{d^2R(\theta')}{d\theta'^2}\right] \right]d\theta'
=0.
\label{3}
\end{equation}
The first term above is just the surface energy, with $\left(R(\theta) +
 d^2R(\theta)/d\theta^2 \right)\,d\theta$ a line element along the domain
 boundary. The second terms is $-\lambda A$, with $\lambda$ a Lagrange
 multiplier. The variational equation reduces to
\begin{equation}
R(\theta)+\frac{d^2R(\theta)}{d\theta^2}=\frac{1}{\lambda}
\left[\sigma(\theta)+\frac{d^2\sigma(\theta)}{d\theta^2}\right],
\label{4}
\end{equation}
with $R(\theta)=\sigma(\theta)/\lambda$ as the solution, modulo an overall
 translation of the domain. { \em The domain shape
 $R(\theta)=\sigma(\theta)/\lambda$ corresponds precisely to the Wulff
 construction}. The somewhat unusual parameterization of two dimensional curves
 thus allows for a straightfoward and analytical determination of the domain
 shape.

The results above have been derived previously \cite{Burtonetal}. However, it
 has not, to our knowledge, been noted that variational derivatives with
respect
 to $R(\theta)$ can be carried out when the anisotropic surface  energy of an
 element of boundary depends on the {\em location} of the element as well as
its
 orientation. One simply replaces $\sigma(\theta)$ by $\sigma(x,y,\theta)$ and
 parameterizes $x$ and $y$ as in Eqs. (\ref{2}). In the remainder of this
Letter
 we describe the consequences of this strategy as applied to a two-dimensional
 domain.

\begin{center}
\underline{ $\kappa$ finite}
\end{center}

We now allow the internal structure of the domain to respond to changes in the
 domain shape. Minimizing ${\rm H}\left[\Theta(x,y)\right]$ with respect to the
 angle $\Theta(x,y)$ leads to the requirement
\begin{mathletters}
\label{5}
\begin{eqnarray}
\nabla^2 \Theta(x,y)&=&0 \\
\left.\kappa\frac{\partial \Theta(x,y)}{\partial n}\right|_{{\rm boundary}} -
 \left.\sigma'(\theta - \Theta(x,y))\right|_{{\rm boundary}}&=&0
\end{eqnarray}
\end{mathletters}
In Eq. (5b) the $\partial/\partial n$ derivative is along the outward normal.
 With no loss of generality we can write the solution of Eq. (5a) as
\begin{equation}
\Theta(x,y)=\frac{1}{i}\left(f(x+iy)-f(x-iy)\right),
\label{6}
\end{equation}
with $f(z)$ an arbitrary analytic function. It is convenient to rescale $x$ and
 $y$ by the mean domain radius $R_0$. The domain boundary in the complex plane
 of a nearly circular domain is, then, the unit circle $z=e^{i \theta}$. Eq.
 (5b) then reduces to
\begin{equation}
\frac{\kappa}{R_0}\left(zf'(z)-1/zf'(1/z)\right) = i\sigma'\left( -i\log
 z+\frac{1}{i}\left(f(z)-f(1/z)\right)\right),
\label{7}
\end{equation}
with $z$ on the unit circle.

If we now go through the same free energy minimization as for $\kappa=\infty$,
 we find instead of Eq. (\ref{4})
\begin{equation}
R(\theta)+\frac{d^2R(\theta)}{d\theta^2}= \sigma_0+{\cal F}(\theta),
\label{8}
\end{equation}
with $\sigma_0$ an isotropic surface tension and
\begin{eqnarray}
{\cal F}(\theta)&=&
 \frac{\kappa}{R_0}\left[z\frac{d}{dz}\left[zf'(z)-\frac{1}{z}
 f'(1/z)\right]-\left[f(z)+f(1/z)\right]\right. \nonumber \\
 &&+ \left. \left.\left(zf(z)\right)^2+\left(
 \frac{1}{z} f(1/z) \right)^2 +\int^z dw\left[ f'(w)^2 - \frac{1}{w^4}
f'(1/w)^2
 \right] \right] \right|_{z=e^{i \theta}}.
\label{9}
\end{eqnarray}
Equations (\ref{8}) and (\ref{9}), the latter relation holding if the domain is
 not too deformed from a perfect circle, are our key results. If we know the
 function $f(z)$ on the unit circle---through Eq. (\ref{7})---then we can
 reconstruct the domain shape with the use of Eqs. (\ref{8}) and (\ref{9}).
Note
 from the discussion following Eq. (\ref{4}) that we can interpret the solution
 $R(\theta)$ of Eq. (\ref{8}) as being proportional to the effective surface
 tension, i. e. a surface tension that incorporates the interior softness of
the
 domain.

Finding the complex function $f(z)$ looks like an intractable problem, as Eq.
 (\ref{7}) is highly nonlinear. We will consider some physically relevant forms
 for $\sigma(\phi)$ to show how $f(z)$ and the domain shape may be found.

\begin{center}
\underline{$\sigma(\phi)=\sigma_0+a_n\cos n \phi$}
\end{center}

The case $n=1$ corresponds to an anisotropic surface energy proportional to
 $\hat{c}\cdot \hat{n}$. The case $n=2$ corresponds to the anisotropy energy of
 a two dimensional nematic.

It can be verified, by direct substitution, that
\begin{equation}
f(z)=\frac{1}{n}\log\left(1-\alpha_nz^n\right)
\label{10}
\end{equation}
is a solution of Eq. (\ref{7}), with
\begin{equation}
\alpha_nR_0^n=\frac{na_nR_0/\kappa}{1+\sqrt{1+\left(na_nR_0/\kappa\right)^2}}.
\label{11}
\end{equation}
When $n=1$ the texture corresponds to the two-dimensional version of a
``virtual
 boojum''\cite{LangSeth,Mermin}---a singularity in the texture that lies
outside
 of the domain.  Imaging of domains of monolayer and near-monolayer films by
 Brewster angle microscopy reveals strong evidence for the existence of this
 structure. \cite{Overbeck,Meunier,LangSethpr}. The distance of the boojum from
 the center of the domain, $R_B$, is given by
\begin{equation}
R_B=R_0\frac{1+\sqrt{1+\left(a_1R_0/\kappa\right)^2}}{a_1R_0/\kappa}.
\label{12}
\end{equation}
If the parameter $\Gamma\equiv\kappa/a_1R_0$ is small compared to one, then
 $R_B$ approaches $R_0$, while in the limit $\Gamma=\infty$ the boojum retreats
 to infinity. The texture is, clearly, highly deformed when $\Gamma<<1$.

The domain shape is found by substituting from Eq. (\ref{10}) into Eq.
 (\ref{9}). Surprisingly, ${\cal F}(\theta)=0$, so $R(\theta)=R_0$. {\em The
 domain shape is thus a perfect circle for $n=1$}. For $n>1$ the exact solution
 for the texture is equivalent to the texture produced by $n$ singularities
 lying outside
of the domain. These singularities are no longer boojums in the
accepted sense of the term \cite{LangSeth,Mermin}. The order parameter angle
 advances by $4\pi/n$ as
one traces a path encircling one of the singularities; an advance of
$4\pi$ characterizes the boojum.  The
singularities for $n>2$ are ``fractionally charged'' in the
topological sense. However, the fractional nature of these
singularities appears to have no consequence for the textural
structure of interest, as the singularities lie outside of the domain. Now
 ${\cal F}(\theta)$ is no longer equal to zero. The domain shape is deformed
 from perfect circularity.

\begin{center}
\underline{ $\sigma(\phi) = \sigma_0+ a_1 \cos \phi + a_2 \cos 2 \phi$}
\end{center}

This is the anisotropic boundary energy believed to be relevant to domains of
 liquid-condensed phase in Langmuir monolayers. It represents the lowest three
 terms in a systematic Fourier expansion of the anisotropic surface energy. For
 $a_2<<a_1$, the function $f(z)$ is given by
\begin{equation}
f(z)=\log\left(1-\alpha z\right) - \frac{a_2}{a_1}\frac{\alpha}{1-\alpha^2}
 \frac{z-\alpha}{1-\alpha z}\int_0^1t^{2
 \alpha^2/(1-\alpha^2)}\left[\frac{zt-\alpha}{1-\alpha zt}\right]\,dt
 +O\left(\left(\frac{a_2}{a_1}\right)^2\right),
\label{13}
\end{equation}
as can be checked by direct substitution. The small admixture of $a_2 \cos 2
 \phi$ in $\sigma(\phi)$ has dramatic effects on the shape. Using Eq.
(\ref{13})
 in Eqs. (\ref{8}) and (\ref{9}), we find that
\begin{itemize}
\item{For $R_0<0.356\sigma_0\kappa/a_1a_2$, the domain has a smooth, nearly
 circular, shape.}
\item{For $R_0>0.356\sigma_0\kappa/a_1a_2$, the domain has a cusp singularity
in
 its shape. For $R_0 \rightarrow \infty$, the cusp angle $\Delta \psi$---the
 difference between the inner angle of the cusp and $\pi$---obeys
\begin{equation}
\Delta \psi(R_0)=5.48 \, \kappa/a_1 R_0 + O\left(1/R_0^2\right).
\label{14}
\end{equation}
Note that the asymptotic variation in the excluded angle is independent of the
 parameter $a_2$. The $a_2 \cos 2 \phi$ term is thus a singular perturbation.
 The domain remains nearly circular outside of the immediate vicinity of the
 cusp.}
\end{itemize}

Large domains will, thus, always have a cusped boundary. Figure \ref{exangle}
 graphs the evolution of the excluded angle of the cusp as a function of the
 radius of a domain in which surface and bulk energy parameters have values
that
  are consistent with the expansion described above. Figure \ref{fishtail}
 depicts a cusped domain as generated by the generalized Wulff construction for
 a specific domain radius. The inner angle of the cusp, $\psi_{{\rm in}}$, is
 indicated in the Figure.

The measurement of cusp angle can be achieved by visual inspection of the
 domain.  In light of the results reported above, a plot of $\Delta \psi$
versus
 $R_0$ should yield the parameter ratio $\kappa/a_1$, and also the combination
 $\sigma_0\kappa/a_1a_2$. A measurement of the cusp angle as a function  of
 domain radius has, in fact, been carried out\cite{Schwartz}. The results will
 be reported elsewhere.

In summary, we have found that deformable domains of materials with an
$XY$-like
 order parameter may have shape singularities, even though their texture is
 perfectly analytic in the domain of definition. It is important to note that
 the material softness plays a key role. If we set the stiffness, $\kappa$, to
 infinity, then the boundary does not have a cusp. The physical origin of our
 cusp is thus strikingly different from that of hard crystalline materials. It
 would be interesting to know whether the results reported here extend to three
 dimensions.

Acknowledgement is due to Dr. Daniel Schwartz and Professor Charles Knobler for
 very useful discussions. J. R. would like to acknowledge the hospitality of
the
 Institute for Theoretical Physics at the Chalmers Institute of Technology and
 the Service de Physique Th\'{e}orique at Saclay, where some of this work was
 carried out.

\begin{figure}
\caption{The excluded angle, $\Delta \psi$, of the cusped domain, defined as
the
 difference beween $\psi_{{\rm in}}$, the inner angle of the cusp (shown in
 Figure 2), and $\pi$. The plot is of the excluded angle versus domain radius
 when two of the parameters controlling surface and bulk energy satisfy $a_2/
 \sigma=0.05$. The domain radius is in units of $\kappa/a_1$.}
\label{exangle}
\end{figure}
\begin{figure}
\caption{the shape of a domain when the parameters controlling surface and bulk
 energy are such that the domain has a cusp. Note that the domain is nearly
 circular, except in the immediate vicinity of the cusp. Barely visible in the
 Figure is the ``swallow tail'' that appears as an appendage, attached to the
 domain at the cusp, when one implements the Wulff construction. This appendage
 is amputated to generate the true domain shape. The virtual boojum lies
outside
 of the domain and near the cusp. In the Figure, $a_2/\sigma=0.25$ and $R_0
 a_1/\kappa=5$. }
\label{fishtail}
\end{figure}

\end{document}